\documentclass[preprintnumbers,amssymb,11pt]{revtex4}
\usepackage{pgfplots}
\usepackage{latexsym,epsfig}
\usepackage{graphicx}
\usepackage{dcolumn}
\usepackage{amsthm,amsmath}

\begin{document}

\title{The role of relativistic many-body theory in probing new physics beyond the standard model via  the electric dipole moments of diamagnetic atoms}

\author{B. K. Sahoo}
\email{bijaya@prl.res.in}
\affiliation{Atomic, Molecular and Optical Physics Division, Physical Research Laboratory, Navrangpura, Ahmedabad 380009, India and\\
State Key Laboratory of Magnetic Resonance and Atomic and Molecular Physics, Wuhan Institute of Physics and Mathematics,
Chinese Academy of Sciences, Wuhan 430071, China}

\author{B. P. Das}
\affiliation{Department of Physics and International Education and Research Center of Science, Tokyo Institute of Technology, 2-12-1 Ookayama Meguro-ku, Tokyo 152-8550, Japan}

\date{Received date; Accepted date}

\begin{abstract}
The observation of electric dipole moments (EDMs) in atomic systems due to parity and time-reversal violating (P,T-odd) interactions can probe new physics beyond the standard model and also provide insights into the matter-antimatter asymmetry in the Universe. The EDMs of open-shell atomic systems are sensitive to the electron EDM and the P,T-odd scalar-pseudoscalar 
(S-PS) semi-leptonic interaction, but the dominant contributions to the EDMs of diamagnetic atoms come from the hadronic and tensor-pseudotensor (T-PT) semi-leptonic interactions. Several diamagnetic atoms like $^{129}$Xe, $^{171}$Yb, $^{199}$Hg, $^{223}$Rn, and $^{225}$Ra are candidates for the  experimental search  for the possible existence of 
EDMs, and among these $^{199}$Hg has yielded the lowest limit till date. The T or CP violating coupling constants of the aforementioned interactions can be extracted from these measurements by combining with 
atomic and nuclear calculations. In this work, we report the calculations of the EDMs of the above atoms by including both the electromagnetic and P,T-odd violating interactions simultaneously. These calculations are performed
by employing relativistic many-body methods based on the random phase approximation (RPA) and the singles and doubles coupled-cluster (CCSD) method starting with the Dirac-Hartree-Fock (DHF) wave function in both cases. The 
differences in the results from both the methods shed light on the importance of the non-core-polarization electron correlation effects that are accounted for by the CCSD method. We also determine electric dipole
polarizabilities of these atoms, which have computational similarities with EDMs and compare them with the available experimental and other theoretical results to assess the accuracy of our calculations.   
\end{abstract}

\maketitle

\section{Introduction}

Attempts to observe an electric dipole moment (EDM) of a non-degenerate system have been steadily gaining ground for the last few decades \cite{graner,parker,hudson,baron,chupp,yamanaka}. 
A non-zero EDM of a non-degenerate system is only possible if the system simultaneously violates parity (P) and time-reversal (T) symmetries \cite{landau,ramsey,fortson}. 
In conformity with the CPT theorem \cite{luders}, T violation implies charge-conjugation and parity (CP) violation. Though there have been direct observations of CP and T 
violations in the neutral K- and B-mesons \cite{christenson,aaij}, but not in other systems. The observed CP and T violations are hadronic in nature, but it is possible that the sources producing leptonic and 
semi-leptonic CP violations could be responsible for the current matter-antimatter asymmetry in the Universe \cite{canetti,dine}. The model dependent analysis of particle physics suggest a range of allowed CP 
violating interactions among elementary particles at different energy scales. The observed CP violations at lower energy scales in the K- and B-mesons have been explained by the most celebrated 
standard model (SM) of elementary particle physics through the Cabibbo-Kobayashi-Maskawa (CKM) matrix. However, this model predicts very small values for the EDM of the electron
and the coupling coefficients for the lepton-hadron CP-violating interactions. Nonetheless, the SM is able to explain many of the observations of  the fundamental properties of elementary 
particles, but it is unable to account for the finiteness of the neutrino masses, the matter-antimatter asymmetry of the Universe, the origin of dark matter and a few other phenomena that are 
of immense interest today. This leads one to believe that the SM is an intermediate manifestation of a more complete theory that needs to be uncovered. Therefore other more sophisticated  extensions of the SM like the multi-Higgs, supersymmetric, left-right symmetric models have been proposed 
\cite{barr,pospelov,fukuyama}. These models could hold the key to some of the profound unresolved questions in particle physics. 

The P and T violating (P,T-odd) interactions producing permanent EDMs of physical systems are very sensitive to beyond the SM (BSM) physics. These interactions correspond to very high energies and are beyond the reach of the present day accelerator facilities like the large hadron collider (LHC). However,  
table-top EDM experiments  are sensitive to CP violating effects at these high energy scales.  Such experiments are currently under way in composite systems like nuclei, atoms, molecules and solids,  
where the interactions are strongly enhanced due to the complex dynamics of many-body effects.  Many high precision measurements in atomic systems 
including EDMs of paramagnetic atoms and polar molecules have been reported \cite{hudson,baron,regan,murthy}. These systems  are sensitive to a limited number of processes 
(e.g. leptonic, hadronic, Higgs, etc.) which arise only in selected particle physics models. In contrast, the sources of EDMs in the diamagnetic atoms can support new physics in multiple 
sectors of a variety of extensions of the SM. The EDMs of diamagnetic atoms arise predominantly from the electron-nucleus (e-N) tensor-pseudotensor (T-PT) interaction and the interaction 
of electrons with the nuclear Schiff moment (NSM) \cite{pospelov,barr1}. The e-N T-PT interaction is the result of the CP violating electron-nucleon (e-n) interactions which arises from
CP violating electron-quark (e-q) interactions at the level of elementary particles that are predicted by the leptoquark models \cite{barr1}. On the other hand, the NSM can exist due to the CP 
violating pion-nucleon-nucleon ($\pi$-n-n) interactions, which in turn have their origins in the CP violating quark-quark (q-q) interactions or the EDMs and chromo-EDMs of quarks that are 
predicted by certain supersymmetric models \cite{barr,pospelov,fukuyama}. In order to obtain precise limits for the coupling constants of these interactions and the EDMs of quarks, 
it is imperative to perform both experiments and relativistic many-body calculations reliably on atoms.

There have been remarkable advances in the past few decades in high precision measurement techniques, particularly those involving the cooling and trapping of atoms and molecules. Some of these have already been
brought to bear on the measurement of EDMs in atomic and molecular systems. A breakthrough in these experiments resulting in the observation of an EDM in the foreseeable future is not beyond the realm of possibility.
In fact, lowering the limits of atomic EDMs ($d_a$) by a few more orders would be desirable as that could discard some of the existing models 
of particle physics. Earlier, spin-exchange pumped masers and a $^3$He co-magnetometer were used by Rosenberry and Chupp to yield an upper limit for the EDM of $^{129}$Xe atom,
$d_a( ^{129}\rm{Xe})=0.7\pm 3.3(\rm{stat})\pm 0.1(\rm{sys})\times 10^{-27}$ e-cm \cite{rosenberry}. Recently, new proposals to measure the EDM of this atom have been made by taking
advantage of its relatively long spin relaxation time \cite{inoue-xe,furukawa-xe,rand,fierlinger,schmidt}. Inoue {\it et al.} \cite{inoue-xe} propose to utilize the nuclear spin oscillator 
technique \cite{yoshimi} for carrying out the measurement of the Larmor precession lower than the above limit. On the other hand, it is anticipated that 
atoms like $^{223}$Rn and $^{225}$Ra will have large enhancements of the EDM signal owing to the large octupole deformed nuclei in these atoms \cite{auerbach}. Motivated by this idea, 
experiments to measure EDMs of these atoms are also in progress \cite{rand}. Even a measurement of the EDM of the radioactive element $^{225}$Ra atom for the first time
has been reported with an upper limit as $|d_a( ^{225}\rm{Ra})|< 5.0 \times 10^{-22}$ e-cm \cite{parker}. However, the lowest EDM limit that has been obtained to date is from 
$^{199}$Hg and it is $|d_a( ^{199}\rm{Hg})| < 7.4 \times 10^{-30}$ e-cm  with 95\% confidence level \cite{graner}.

As mentioned earlier, only the combined results of the atomic and nuclear calculations with the measured values of the EDMs of atoms are useful to constrain various CP-odd coupling constants.
A number of EDM calculations have been performed on atoms that are of experimental interest, i.e. $^{129}$Xe, $^{171}$Yb, $^{199}$Hg, $^{223}$Rn, and $^{225}$Ra by employing a variety of 
relativistic atomic many-body methods (see for more details Ref. \cite{yamanaka}). In our recent works \cite{yashpal-xe,bks-rn,yashpal-hg,yashpal-ra,bks-yb}, we show that the results from
the relativistic coupled-cluster (RCC) method differ by about 20-50\% from the previous calculations on the aforementioned atoms that are obtained by other approximate all-order methods like the random 
phase approximation (RPA) \cite{martensson,dzuba02,latha-rpa}, multi-configuration Dirac-Fock (MCDF) method \cite{angom,jacek}, a combined configuration interaction (CI) method with approximate 
many-body perturbation theory (MBPT ) \cite{dzuba09}, and also another perturbed RCC (PRCC) method \cite{latha}. The differences between these methods have been extensively discussed by us
in our earlier works \cite{yashpal-xe,bks-rn,yashpal-hg,yashpal-ra,bks-yb} and the details can be found in a recent review \cite{yamanaka}. The RPA method accounts for only the mean-field and 
the core-polarization effects to all-order. The coupled-cluster (CC) method, which has been referred to as the gold standard for many-body theory \cite{bishopbook,bartlett,lindgren} has been applied 
to nuclear \cite{hagen}, atomic \cite{deyonker,nataraj}, molecular \cite{crawford,prasana} and condensed matter systems \cite{bishop1,bishop2}. Some of the notable features of this method is that 
even the truncated (R)CC methods at a given level of excitation is equivalent to including correlation effects to all-orders in the residual interaction
\cite{bishopbook,bartlett,crawford}, owing to the exponential {\it ansatz} in expressing  the wave function, it is size-extensive and size-consistent  \cite{bishopbook,bartlett} and amenable to high performance
computational techniques  \cite{bartlett,crawford,kallay1,kallay2,hirata}. It is also necessary 
to use relativistic theory for studying EDMs in atomic systems as EDMs are prominent only in heavy atomic systems in which relativistic effects are strong. Also, contributions from 
the electron EDM to the atomic EDM can only be explained by a relativistic theory \cite{sandars1,sandars2,sandars3,das}. 

In this paper, we present our results for $d_a$ for  $^{129}$Xe, $^{171}$Yb, $^{199}$Hg, $^{223}$Rn, and $^{225}$Ra atoms due to the T-PT and NSM PT-odd interactions by employing the 
RCC method. These values are then compared with their corresponding values from the DHF and RPA methods in order to highlight the role of electron correlations for their 
determination and find the importance of the non-core-polarization that are neglected by RPA. We also carry out calculations of the electric dipole polarizabilities
($\alpha_d$) of the above atoms using the DHF, RPA and RCC methods and compare them with the available experimental values and other theoretical calculations. The purpose of performing these 
calculations is the similarity of the mathematical expressions of $d_a$ and $\alpha_d$, thus the reliability of the results of the calculations of $d_a$ can be judged to some extent by comparing the 
results of $\alpha_d$ with their measured values wherever available, or else comparing with other calculations that are reported by applying sophisticated many-body methods.  
 
The rest of the paper is organized as follows: In the next section, we briefly mention about the theory of atomic EDMs and present  the T-PT and NSM interaction Hamiltonians used for the EDM
calculations. Then, we describe our many-body methods and procedures for obtaining atomic wave functions at various levels of approximations. This is followed by discussions on our results and 
comparison of these values with the previously performed calculations. Unless stated otherwise, we use atomic units (a.u.) throughout this paper.

\section{Dominant sources of P,T-odd interactions}

We start from the e-n P,T-odd interactions that lead to the dominant P,T-odd interactions in diamagnetic atoms. To arrive at the T-PT e-N interaction in an atom, we write 
the P,T-odd Lagrangian for the e-n interaction as \cite{pospelov}
\begin{eqnarray}
\mathcal{L}_{e-n}^{PT} &=& C_T^{e-n} \varepsilon_{\mu \nu \alpha \beta} \bar{\psi}_e \sigma^{\mu \nu} \psi_e  
\bar{\psi}_n \sigma^{\alpha \beta} \psi_n + C_P^{e-n}  \bar{\psi}_e  \psi_e \ \bar{\psi}_n i \gamma_5 \psi_n,
\end{eqnarray}
where $\varepsilon_{\mu \nu \alpha \beta}$ is the Levi-Civita symbol and $\sigma_{\mu \nu} = \frac{i}{2}[\gamma_\mu, \gamma_\nu]$ with the $\gamma$'s being the Dirac matrices. $C_T^{e-n}$ and 
$C_P^{e-n}$ are the tensor-pseudotensor (T-PT) and scalar-pseudoscalar (S-PS) e-n interaction coupling constants respectively. Here $\psi_n$ and $\psi_e$ represent the field operators for 
the nucleon and electron respectively. This gives the T-PT e-N interaction Hamiltonian for atoms as \cite{yamanaka}
\begin{eqnarray}
H_{e-N}^{TPT} &=& i \sqrt{2} G_F C_T \sum_e \mbox{\boldmath $\sigma_N \cdot \gamma$} \rho_N(r) ,
\label{tpteq}
\end{eqnarray}
where $G_F$ is the Fermi constant, $C_T$ is the T-PT coupling constant, {\boldmath$\sigma_N$}$=\langle \sigma_N \rangle {\bf I}/I$ is the Pauli spinor of the nucleus with spin $I$, and
$\rho_N(r)$ is the nuclear density.

Similarly, the Lagrangian for the other dominant P,T-odd $\pi$-n-n interaction is given by \cite{pospelov}
\begin{eqnarray}
\mathcal{L}^{\pi n n}_{e-n} &=& \bar{g}_{\pi n n}^{(0)} \bar{\psi}_n \tau^i \psi_n \pi^i + \bar{g}_{\pi n n}^{(1)} \bar{\psi}_n \psi_n \pi^0  + \bar{g}_{\pi n n}^{(2)} \big ( \bar{\psi}_n \tau^i \psi_n \pi^i - 
3 \bar{\psi}_n \tau^3 \psi_n \pi^0 \big )
\end{eqnarray}
where the couplings $\bar{g}_{\pi n n}^{(i)}$ with the superscript $i=$ 1, 2, 3 represent the isospin components. This yields the corresponding e-N interaction Hamiltonian as \cite{yamanaka}
 \begin{eqnarray}
  H_{e-N}^{NSM}= \frac{3{\bf S.r}}{B_4} \rho_N(r),
 \label{nsmeq}
 \end{eqnarray}
where ${\bf S}=S \frac{{\bf I}}{I}$ is the nuclear Schiff moment (NSM) and $B_4=\int_0^{\infty} dr r^4 \rho_N(r)$. The magnitude of NSM $S$ is given by \cite{engel,haxton}
\begin{eqnarray}
S = g_{\pi n n} \times (a_0 \bar{g}_{\pi n n}^{(0)} + a_1 \bar{g}_{\pi n n}^{(1)} + a_2 \bar{g}_{\pi n n}^{(2)}),
\end{eqnarray}
where $g_{\pi nn} \simeq 13.5$ is the CP-even $\pi$-n-n coupling constant and $a_i$ are the polarization parameters of the nuclear charge distribution that can be computed to a reasonable accuracy using
the Skyrme effective interactions in the Hartree-Fock-Bogoliubov mean-field method \cite{engel}. These couplings are related to the chromo-EDMs of up-quark ($\tilde{d}_u$) and down-quark ($\tilde{d}_d$) 
as $\bar{g}_{\pi n n}^{(1)} \approx 2 \times  10^{-12} \times (\tilde{d}_u - \tilde{d}_d)$ \cite{pospelov,pospelov1} and $\bar{g}_{\pi n n}^{(0)}/ \bar{g}_{\pi n n}^{(1)} \approx 0.2 \times (\tilde{d}_u 
+ \tilde{d}_d)/(\tilde{d}_u -\tilde{d}_d)$ \cite{pospelov,dekens}, where $\tilde{d}_u$ and $\tilde{d}_d$ are scaled to $10^{-26}$ e-cm. Also, it yields a relation with the quantum chromodynamics (QCD) 
parameter ($\bar{\theta}$) by $|\bar{g}_{\pi n n}^{(1)}|=0.018(7) \bar{\theta}$ \cite{dekens}. From the nuclear calculations, one can obtain $S \simeq (1.9d_n+0.2d_p)$ fm$^2$ \cite{dmitriev}. Thus, it is
necessary to obtain accurate values of $C_T$ and $S$ by combining atomic calculations with the experimental EDM result to extract the magnitudes of the coupling constant $C_T^{e-n}$  and the other fundamental 
CP violating parameters mentioned above in a reliable manner.

\section{Relativistic Atomic Many-body Methods}

The EDM of the ground state $|\Psi_0 \rangle$ in an atom is given by
\begin{eqnarray}
 d_a = \frac{\langle \Psi_0 | D | \Psi_0 \rangle}{\langle \Psi_0 | \Psi_0 \rangle },
 \label{edmeq}
\end{eqnarray}
where $D$ is the electric dipole moment operator. $D$ is an odd-parity operator, so its expectation value will be zero if we consider only the electromagnetic interactions in the 
atoms. To obtain a non-zero EDM, the state $|\Psi_0 \rangle$ must be obtained after including both the electromagnetic and weak interactions in the atomic systems. The dominant  electromagnetic
interactions in an atom arise due to the exchange of a single photon, and therefore we have performed our calculations in the present work in this approximation. These calculations exploit the spherical symmetry of the atoms.   
It would be appropriate to determine the atomic states considering first the dominant electromagnetic interactions alone and later
treat the weak interaction of interest as a first-order perturbation. Consideration of both the electromagnetic and P,T-odd interactions simultaneously will mix orbitals of different parities and lead to a large 
configuration space for determining atomic wave functions, which require large scale computations. It is possible to reduce these computations substantially via an appropriate group symmetry for cases where the molecular orbitals are described using Cartesian coordinates. On the other hand, atomic orbitals due to the dominant electromagnetic interactions alone are described well by spherical coordinates leading to parity as a good
quantum number in this case, thereby resulting in a reduction in the computational effort in performing the atomic calculations. P,T-odd interactions that have been considered in this work can be treated as first-order perturbations as their strengths are very weak. 

To evaluate the atomic state functions due to the electromagnetic interactions, we consider the Dirac-Coulomb (DC) Hamiltonian which is given by
\begin{eqnarray}
H^{DC} &=& \sum_i \Lambda_i^+  \left [ c\mbox{\boldmath$\alpha$}_i\cdot \textbf{p}_i+(\beta_i -1)c^2 + V_N(r_i) \right ] \Lambda_i^+
+ \sum_{i,j\ge i} \Lambda_i^+ \Lambda_j^+ \frac{1}{r_{ij}} \Lambda_i^+ \Lambda_j^+ , 
\end{eqnarray}
where $\mbox{\boldmath$\alpha$}$ and $\beta$ are the usual Dirac matrices, $c$ is the speed of light, $V_N(r)$ represents for the nuclear potential and $\Lambda^+$ operators for the respective 
orbitals correspond to the projection operators on to positive energy levels of the Dirac orbitals. We evaluate the nuclear potential considering the Fermi-charge distribution defined by
\begin{equation}
\rho_N(r)=\frac{\rho_{0}}{1+e^{(r-b)/a}},
\end{equation}
for the normalization factor $\rho_0$, the half-charge radius $b$ and $a= 2.3/4(ln3)$ are related to the skin thickness. We have used $a= 2.3/4(ln3)$ and $b$ is determined using the relation 
\begin{eqnarray}
b&=& \sqrt{\frac {5}{3} r_{rms}^2 - \frac {7}{3} a^2 \pi^2}
\end{eqnarray}
with the root mean square (rms) charge radius of the nucleus determined using the formula
\begin{eqnarray}
 r_{rms} = 0.836 A^{1/3} + 0.570
\end{eqnarray}
in $fm$ for the atomic mass $A$.

\begin{table}[h]
\caption{Results of $\alpha_d$ values in $e a_0^3$ and  ${\cal R}$ due to T-PT (in $10^{-20} \langle \sigma\rangle |e|cm$) and NSM (in $10^{-17}[1/|e|fm^3]|e|cm$) P,T-odd
interactions for the considered diamagnetic atoms from DHF, RPA and CCSD methods. The $\alpha_d$ values are compared with available experimental values wherever available (else 
we mention NA for not available) and other theoretical calculations from different many-body methods including a variant of CCSD methods that are reported in the 
literature.}
\begin{tabular}{lcccccc}
\hline \hline \\
 Method   & $^{129}$Xe &  $^{171}$Yb & $^{199}$Hg & $^{223}$Rn & $^{225}$Ra \\ 
\hline \\
          \multicolumn{6}{c}{$\alpha_d$ value}   \\
DHF  &  26.87 & 124.51  & 40.95  & 34.42 & 204.13  \\       
RPA  &  26.98 & 179.51  & 44.98  & 35.00 & 296.85   \\ 
CCSD &  28.13  & 134.82  & 34.51  & 36.60 & 228.68   \\
 ($ \Delta$B) & (0.04) & (0.49) & ($-0.01$) & (0.09) & (0.19) \\
 Final & 28.1(5)  & 135(5) & 34.5(8) & 36.6(5) &  229(15) \\ 
           &      &      &     &      &      \\
 Others    &      &      &     &      &     \\ 
 RPA  & 27.7 \cite{latha-rpa} & 179 \cite{dzuba02}, 176.16 \cite{latha-rpa} & 44.92 \cite{dzuba02}   & 35.00 \cite{dzuba09} & 291.4 \cite{latha-rpa} \\
 CI$+$MBPT &    &   & 229.9 \cite{dzuba09}  \\
 CCSD(T)   &    &   &  34.42 \cite{seth}  &    & 242.8 \cite{borschevsky} \\ 
 PRCC   &   26.432 \cite{sidh2} &   &  33.294 \cite{latha}  & 35.391 \cite{sidh2} &  \\
Experiment & 27.815(27) \cite{hohm-xe} & 142(36) \cite{miller}  & 33.91(34) \cite{goebel} & NA & NA \\
\hline \\
          \multicolumn{6}{c}{${\cal R}$ due to T-PT}   \\ 
DHF  &  0.45 & $-0.71$  & $-2.39$  & 4.48 & $-3.46$  \\       
RPA  &  0.56 & $-3.39$  & $-5.89$  & 5.40 & $-16.66$  \\
CCSD &  0.47  & $-2.03$  & $-3.17$ & 4.43  & $-9.81$   \\
 ($ \Delta$B) & ($-0.01$) & (0.02) & (0.03) & ($-0.02$) & (0.06) \\
 Final & 0.47(3) & $-2.0(3)$ & $-3.2(5)$ & 4.4(5) & $-9(1)$ \\ 
           &      &      &     &      &      \\
 Others    &      &      &     &      &     \\ 
 RPA  & 0.564 \cite{latha-rpa}, 0.57 \cite{dzuba09} & $-3.4$ \cite{dzuba02} & & 5.6 \cite{dzuba09}   & $-16.59$ \cite{latha-rpa} \\ 
 CI$+$MBPT &      &      &  $-5.1$ \cite{dzuba09}   &  &  $-18$ \cite{dzuba09} \\ 
 MCDF      &      & $-4.84$ \cite{jacek} &  $-4.3$ \cite{latha} &  & \\
 PRCC      &      &      &  \\
 \hline \\
          \multicolumn{6}{c}{${\cal R}$ due to NSM}   \\ 
DHF  & 0.29 & $-0.42$  & $-1.20$  & 2.46 & $-1.85$  \\       
RPA  & 0.37 & $-1.91$  & $-2.94$  & 3.31 & $-8.12$   \\
CCSD & 0.33  & $-1.49$ & $-1.76$  & 2.77 & $-6.13$   \\
 ($\Delta$B) & ($\sim 0.0$) & (0.02) & (0.04) & ($-0.03$) & (0.06) \\
 Final & 0.33(2) & $-1.5(3)$ & $-1.8(3)$ & 2.8(3) & $-6.1(5)$ \\  
            &      &      &     &      &      \\
 Others    &      &      &     &      &     \\ 
 RPA  & 0.38 \cite{dzuba09} & $-1.903$ \cite{latha-rpa} &  & 3.3 \cite{dzuba09} &  \\ 
 MCDF &                & $-2.51$ \cite{angom}, $-2.15$ \cite{radziute} & $-2.22$ \cite{jacek} &  &  \\
 CI$+$MBPT       &  &  $-2.12$ \cite{dzuba09}  & 2.6 \cite{dzuba09}  & & $-8.8$ \cite{dzuba09}  \\
 PRCC            &  &   & $-2.46$ \cite{latha}  &  &  \\
\hline \hline\\
\end{tabular}
\label{tab1}
\end{table}

We have also included the frequency-independent Breit interaction, which is the leading relativistic correction to the two-body Coulomb interaction in our present work. This interaction is given by 
\begin{eqnarray}
V_B(r_{ij})=-\frac{1}{2r_{ij}}\{\mbox{\boldmath$\alpha$}_i\cdot \mbox{\boldmath$\alpha$}_j+
(\mbox{\boldmath$\alpha$}_i\cdot\bf{\hat{r}_{ij}})(\mbox{\boldmath$\alpha$}_j\cdot\bf{\hat{r}_{ij}}) \} .
\end{eqnarray}

As mentioned earlier, we only consider first-order terms in the P,T-odd weak interactions. The total atomic Hamiltonian can, therefore, be written as
\begin{eqnarray}
 H = H^{at} + \lambda H^{PT},
\end{eqnarray}
where $H^{at}$ represents  the atomic Hamiltonian consisting of the DC Hamiltonian alone or  with the Breit interaction added to it and $\lambda H^{PT}$ corresponds to either of the P,T-odd 
Hamiltonians given by Eqs. (\ref{tpteq}) and (\ref{nsmeq}). Here $\lambda$ can be $S$ or $C_T$ for the P,T-odd Hamiltonians. In this approach, the atomic wave function can be expanded as
\begin{eqnarray}
 |\Psi_0 \rangle \simeq |\Psi_0^{(0)} \rangle + \lambda |\Psi_0^{(1)} \rangle  
\end{eqnarray}
keeping terms up to first-order in $\lambda$, where $| \Psi_0^{(0)} \rangle$ and $|\Psi_0^{(1)} \rangle$ are the wave functions of $H^{at}$ and its first-order correction due to one of the P,T-odd 
interaction Hamiltonians, respectively. Hence Eq. (\ref{edmeq}) can be expressed as
\begin{eqnarray}
 d_a &=& 2 \lambda \frac{\langle \Psi_0^{(0)}|D|\Psi_0^{(1)} \rangle}{\langle \Psi_0^{(0)}|\Psi_0^{(0)} \rangle}.
 \label{eqed}
\end{eqnarray}
The actual quantity that is calculated is
\begin{eqnarray}
 {\cal R} = {d_a} / {\lambda} &=& 2 \frac{\langle \Psi_0^{(0)}|D|\Psi_0^{(1)} \rangle}{\langle \Psi_0^{(0)}|\Psi_0^{(0)} \rangle}
\label{eqptt}
 \end{eqnarray}
so that it can be combined with the experimentally measured $d_a$ values to extract $\lambda$.

\begin{table}[h]
\caption{Contributions to $\alpha_d$ values in $e a_0^3$ and ${\cal R}$ due to T-PT (in $10^{-20} \langle \sigma\rangle |e|cm$) and NSM (in $10^{-17}[1/|e|fm^3]|e|cm$) P,T-odd
interactions for the considered diamagnetic atoms from different RCC and their h.c. terms of the CCSD method. Here $\tilde{D}_{ob}$ represents effective one-body operators of
$\tilde{D}$, which is a non-terminating series in the CCSD method but has been evaluated adopting self-consistent approach.}
\begin{tabular}{lccccc}
\hline \hline \\
 RCC terms   & $^{129}$Xe &  $^{171}$Yb & $^{199}$Hg & $^{223}$Rn & $^{225}$Ra \\ 
\hline \\
          \multicolumn{6}{c}{$\alpha_d$ value}   \\
$\tilde{D}_{ob}T_1^{(1)}+$h.c.                & 26.672    & 125.217   &  33.034   & 34.748   & 210.845  \\
$T_1^{(0)\dagger}\tilde{D}_{ob}T_2^{(1)}+$h.c. & 0.084     & 0.757     &  0.044    & 0.085     & 1.561 \\
$T_2^{(0)\dagger}\tilde{D}_{ob}T_2^{(1)}+$h.c. & 1.370     & 8.842     &  1.429    & 1.772     & 16.276 \\
\hline \\
          \multicolumn{6}{c}{${\cal R}$ due to T-PT}   \\ 
$\tilde{D}_{ob}T_1^{(1)}+$h.c.                &  0.482    & $-2.009$ & $-3.138$  & 4.551     & $-9.637$ \\
$T_1^{(0)\dagger}\tilde{D}_{ob}T_2^{(1)}+$h.c. &  $-0.005$ & $-0.021$ & $-0.027$  & $-0.054$  & $-0.065$ \\
$T_2^{(0)\dagger}\tilde{D}_{ob}T_2^{(1)}+$h.c. &  $-0.004$ & 0.003    & 0.022     & $-0.073$  & $-0.108$ \\
\hline \\
          \multicolumn{6}{c}{${\cal R}$ due to NSM}   \\
$\tilde{D}_{ob}T_1^{(1)}+$h.c.                & 0.325      & $-1.417$  &  $-1.743$ & 2.745    & $-5.780$ \\
$T_1^{(0)\dagger}\tilde{D}_{ob}T_2^{(1)}+$h.c. & $\sim 0.0$  &  0.014  &  0.008  & $-0.004$ & 0.035 \\
$T_2^{(0)\dagger}\tilde{D}_{ob}T_2^{(1)}+$h.c. & 0.005     & $-0.088$  & $-0.022$  &  0.029   & $-0.389$ \\
\hline \hline \\
\end{tabular} 
\label{tab2}
\end{table}

The first-order perturbed wave function $|\Psi^{(1)} \rangle$ can be calculated by two different approaches. In the sum-over-states approach, we use the expression for the first-order perturbed wave function
\begin{eqnarray}
 |\Psi_0^{(1)} \rangle = \sum_{I \ne 0} |\Psi_I^{(0)} \rangle \frac{\langle \Psi_I^{(0)} | H^{PT}|\Psi_0^{(0)}\rangle}{E_0^{(0)} - E_I^{(0)}}
\end{eqnarray}
where $|\Psi_I^{(0)} \rangle$s are the eigen states other than $|\Psi_0^{(0)}\rangle$ of $H^{at}$ with the energies $E_I^{(0)}$ and $E_0^{(0)}$ corresponds to the ground state energy. 
This leads to the following expression for ${\cal R}$ 
\begin{eqnarray}
  {\cal R} &=& 2 \sum_{I \ne 0} \frac{\langle \Psi_0^{(0)}|D|\Psi_I^{(0)} \rangle\langle \Psi_I^{(0)} | H^{PT}|\Psi_0^{(0)}\rangle }{\langle \Psi_0^{(0)}|\Psi_0^{(0)} 
  \rangle (E_0^{(0)} - E_I^{(0)}) } .
\end{eqnarray}
Replacing $H^{PT}$ by $D$ operator in the above expression will give $\alpha_d$ value which can be measured precisely for many atomic systems. This is the reason why electric 
dipole polarizabilities are calculated and compared with the available experimental data and used as a metric to ascertain the reliability  of the calculated value of ${\cal R}$.
The advantage
of adopting the sum-over-states approach is that one has to calculate only the dominant low-lying state contributions  and the uncertainties  can be reduced
 by using experimental
energies. However, this method cannot evaluate correlation contributions from the core and the continuum states which can be significant for heavy atomic systems. 
 The most appropriate approach to obtain the first-order perturbed wave function involves solving the following inhomogeneous equation
\begin{eqnarray}
(H^{at}-E_0^{(0)}) |\Psi_0^{(1)} \rangle &=& (E_0^{(1)}- H^{PT}) |\Psi_0^{(0)} = - H^{PT} |\Psi_0^{(0)} \rangle ,
\label{eq4}
\end{eqnarray}
where the first-order perturbed energy $E_0^{(1)}$ vanishes since $H^{PT}$ has odd parity. It is well known that the determination of $|\Psi_0^{(0)} \rangle$ itself accurately in heavy 
atomic systems is challenging  because of the of two-body Coulomb and Breit interactions. So the choice of a relativistic many-body method for the evaluation of  $|\Psi_0^{(0)} \rangle$  and $|\Psi_0^{(1)} \rangle$ has to be made judiciously.   

We discuss here two variants of all-order relativistic many-body methods, RPA and RCC, to determine ${\cal R}$ for the aforementioned atoms that are of experimental interest, i.e. $^{129}$Xe, $^{171}$Yb, 
$^{199}$Hg, $^{223}$Rn, and $^{225}$Ra. Earlier, relatively simple many-body methods such as the relativistic third-order many-body perturbation theory (MBPT(3) method) and the relativistic 
RPA were employed to determine these quantities in $^{129}$Xe and $^{223}$Rn \cite{martensson,dzuba02}. These methods are not sufficiently powerful for performing rigorous calculations of ${\cal R}$  for atoms such as 
$^{171}$Yb, $^{199}$Hg and $^{225}$Ra, where pair-correlation effects contribute significantly. So the combined CI$+$MBPT method has been employed to evaluate ${\cal R}$ values in 
these atoms \cite{dzuba09}. In this hybrid method, the single particle orbitals are determined using a $V^{N_c-2}$ potential, where $N_c$ is the total number of electrons and the electron 
correlation effects are accounted for by dividing the electrons into valence and core electrons. 

To demonstrate the role electron correlation effects in the evaluation of ${\cal R}$ values, we first evaluate the DHF contribution using the expression 
\begin{eqnarray}
 {\cal R} & =& 2 \sum_{I \ne 0} \frac{  \langle \Phi_0| D| \Phi_I \rangle \langle \Phi_I | H^{PT} | \Phi_0 \rangle} { E_0^{DHF} - E_I^{DHF}} ,
\end{eqnarray}
where $|\Phi_0 \rangle$ and $| \Phi_I \rangle$ are the DHF wave functions of the ground and possible excited states of an atom and $E_{0,I}^{DHF}$ are their respective energies. Since $D$ is 
an one-body operator $| \Phi_I \rangle$ represents singly excited determinants with respect to $|\Phi_0 \rangle$.

 By perturbing the DHF orbitals to first-order and adopting a self-consistent procedure, we can express the first-order perturbed wave function due to the P,T-odd interaction as
\begin{eqnarray}
 |\Psi_0^{(1)} \rangle & \approx & \Omega_{RPA}^{(1)} | \Phi_0 \rangle ,
\end{eqnarray}
where we define
\begin{eqnarray}
\Omega_{RPA}^{(1)} &=&  \sum_k^{\infty} \sum_{p,a} \Omega_a^{p(k, 1)} \nonumber \\
    &=& \sum_{\beta=1}^{\infty} \sum_{pq,ab} \left \{ \frac{[\langle pb | V(r_{ij})  | aq \rangle 
- \langle pb | V(r_{ij}) | qa \rangle] \Omega_b^{q(\beta-1,1)} } {\epsilon_a - \epsilon_p} \right \} \nonumber \\
 && +\sum_{\beta=1}^{\infty} \sum_{pq,ab} \left \{ \frac{ \Omega_b^{q{(\beta-1,1)}^{\dagger}}[\langle pq | V(r_{ij}) | ab \rangle - \langle pq | V(r_{ij}) | ba \rangle] 
}{\epsilon_a-\epsilon_p} \right  \} ,
\label{eq27}
\end{eqnarray} 
with the initial guess $\Omega_a^{p(0, 1)} |\Phi_0 \rangle = \frac{\langle \Phi_a^p | H^{PT} | \Phi_0 \rangle } {\epsilon_a - \epsilon_p}$, for the singly excited determinants $|\Phi_a^p \rangle$
with subscript $a$ and superscript $p$ representing single excitation configurations by replacing an occupied orbital $a$ from $|\Phi_0 \rangle$ by a virtual orbital $p$, $\epsilon$ stands for 
the single particle orbital energies and superscript $k$ represents for orders of two-body interactions taken into account, which goes to infinity in this case.  Adopting this procedure, we can 
evaluate ${\cal R}$ by
\begin{eqnarray}
 {\cal R}  & \approx & 2 \langle \Phi_0| D \Omega_{RPA}^{(1)} |\Phi_0 \rangle .
\end{eqnarray} 
It can be shown from the above formulation that the RPA method subsumes a certain class of single excitations representing  the polarization of the core by the P,T-odd interaction  to all-orders. This would be the dominant correlation effect
for the inert gas atoms, where the pair-correlation effects are not very significant in the all-order perturbation calculations and this is evident from the RCC calculations.

 Taking $|\Phi_0 \rangle$ as the reference state, $|\Psi_0^{(0)} \rangle$ is expressed in the (R)CC method as \cite{cizek} 
\begin{eqnarray}
 |\Psi_0^{(0)} \rangle = e^{T^{(0)}} |\Phi_0 \rangle,
 \label{eqt0}
\end{eqnarray}
where $T^{(0)}$ is the even parity RCC excitation operator due to the residual Coulomb interaction; i.e. the difference of the two-body Coulomb and the DHF potential energies. In the particle-hole 
excitation formalism, we express $T^{(0)}=\sum_{I=1}^{N_c} t_I^{(0)} C_I^{+}$, $t_I^{(0)}$ are the amplitudes of the excitations and $C_I^{+}$ stands for a string of annihilation-creation operators 
corresponding to a general particle-hole excitation. The equation for the ground state of $H^{at}$ is given by
\begin{eqnarray}
 H^{at} |\Psi_0^{(0)} \rangle &=& E_0^{(0)} |\Psi_0^{(0)} \rangle
 \label{eqm0}
\end{eqnarray}
with the exact ground state energy $E_0^{(0)}$. The equations for the cluster amplitudes $T^{(0)}$ are obtained by using Eq. (\ref{eqt0}) and projecting Eq. (\ref{eqm0}) on the bra state
$\langle \Phi_0| C_I^{-}e^{-T^{(0)}}$ as 
\begin{eqnarray}
\langle \Phi_0| C_I^{-} \overline{H}_N^{at}|\Phi_0\rangle=0,  \label{eq36} 
\end{eqnarray}
where the de-excitation operators $C_I^{-}$ are the hermitian conjugate (h.c.) of $C_I^{+}$ and $H_N^{at} = H^{at}-\langle \Phi_0 | H^{at} | \Phi_0 \rangle$ is the normal order Hamiltonian. We use the 
notation $\overline{O}=e^{-T}Oe^T=(Oe^T)_{l}$ through out the paper for a general operator $O$, where the subscript $l$ stands for the linked terms \cite{bartlett}. For one-body and two-body operators 
$O$, $\overline{O}$ terminates naturally \cite{bartlett,crawford}.

The inclusion of a weak P,T-odd interaction Hamiltonian or the electric dipole operator $D$, denoted by $H_{\lambda}$ now onwards, will modify the ground state wave function 
which can be written as 
\begin{eqnarray}
 |\Psi_0 \rangle = e^T |\Phi_0 \rangle=e^{T^{(0)}+\lambda T^{(1)}} |\Phi_0 \rangle,
\end{eqnarray}
where the effect of the perturbation is represented by $T^{(1)}=\sum_{I=1}^{N_c} t_I^{(1)} C_I^{+}$ with the amplitudes $t_I^{(1)}$, which includes one-order of the weak odd-parity
perturbation of interest and all-orders of the residual interaction. Here $\lambda$ represents the strength of the coupling coefficient of a given P,T-odd interaction or the 
electric field for the evaluation of $d_a$ and $\alpha_d$, respectively.  The first-order perturbed wave function can be identified as
\begin{eqnarray}
 |\Psi_0^{(1)} \rangle = e^{T^{(0)}} T^{(1)} |\Phi_0 \rangle.
 \label{eqt1}
\end{eqnarray}
The equation for the amplitudes of $T^{(1)}$ can be obtained by solving \cite{yashpal-xe,bks-rn,yashpal-hg,yashpal-ra,bks-yb}
\begin{eqnarray}
\langle \Phi_0|C_I^{-} \left [ \overline{H}_N^{at}T^{(1)} + \overline{H}_{\lambda}\right ] |\Phi_0\rangle =0 . \label{eq37} 
\end{eqnarray}
For the calculations of $|\Psi_0^{(0)} \rangle$ and $|\Psi_0^{(1)} \rangle$, we consider singles (one particle-one hole) and doubles (two particle-two hole) excitations 
in the RCC theory (CCSD method) by restricting to $I=1,2$ in the amplitude equations. From the point of view of computational efficiency, we construct the 
intermediate diagrams of $\overline{H}_N^{at}$ by dividing it into effective one-body and two-body intermediate operators as described in detail in Refs. \cite{yamanaka,yashpal1,
yashpal2}. 

Following the above expression and expanding $|\Psi_0 \rangle$, we can evaluate $d_a$ and $\alpha_d$, commonly denoted here as $X$, for the ground state of a closed-shell atom by \cite{yashpal-hg,bartlett,bishopbook,cizek}
\begin{eqnarray}
X &\equiv& \frac{\langle \Psi_0 | D | \Psi_0 \rangle}{\langle \Psi_0 | \Psi_0 \rangle} =  \langle \Psi_0 | D | \Psi_0 \rangle_l = \lambda \langle \Psi_0^{(1)} | D | \Psi_0^{(0)} \rangle + \langle \Psi_0^{(0)} | D | \Psi_0^{(1)} \rangle \nonumber \\ 
&=&  \lambda \langle\Phi_0 | e^{T^{(0) \dagger}} D e^{T^{(0)}} T^{(1)} + T^{(1)\dagger} e^{T^{(0)\dagger}} D e^{T^{(0)}} | \Phi_0 \rangle_l \nonumber \\
                  &=& 2 \lambda \langle\Phi_0 | \tilde{D}^{(0)} T^{(1)} | \Phi_0 \rangle_l ,  \label{eqccx}
\end{eqnarray}
where $\tilde{D}^{(0)}= e^{T^{(0)\dagger}} D e^{T^{(0)}}$ is a non-terminating series. The terms in the series can be computed uniquely by ensuring that they
do not repeat. This is achieved  by first considering the linear terms and contracting $D$ that appears in $\overline{D}^{(0)}$  with a  $T^{(0)}$ or $T^{\dagger (0)}$ 
operator. Next, the contributions from the non-linear terms are considered by and contractions are carried out with other $T^{(0)}$ and $T^{\dagger (0)}$ operators till self-consistent results are obtained.

 In order to estimate the uncertainty in the calculations due to the neglected triple excitations in the CCSD method, we define an excitation operator using perturbation theory in the RCC framework in the 
following manner
\begin{eqnarray}
 T_3^{(0),pert}= \frac{1}{3!}\sum_{abc,pqr}  \frac{ ( \overline{H}_N^{at} T_2^{(0)})_{abc}^{pqr} }{\epsilon_a  + \epsilon_b+\epsilon_c-\epsilon_p -\epsilon_q -\epsilon_r} ,
 \label{eq30}
\end{eqnarray}
where $a,b,c$ and $p,q,r$ subscripts denote for occupied and unoccupied orbitals, respectively. From the differences between the results of the CCSD method and the 
calculations carried out with the inclusion of the above operator as a part of $T^{(0)}$, we find an order of magnitude estimate of the contributions of the triple excitations. It is pertinent to note 
that the triple excitation contributions coming through the $T^{(1)}$ RCC operators will be extremely small.  We have estimated the above corrections from the singles 
equations of the CCSD method for reasons of computational simplicity in order get an idea about the sizes of the errors in our calculations  of the properties that we have calculated in the present work. 

\section{Results and Discussion}

In Table \ref{tab1}, we present $\alpha_d$ values and ${\cal R}$ due to both the T-PT and NSM interactions from our DHF, RPA and CCSD calculations for all the atoms that have been considered in this study.
The polarizability values are compared with the available experimental results and other calculations employing a variety of relativistic many-body methods as mentioned in 
the same table. By comparing the RPA and CCSD results with the DHF values, it can be seen that the correlation trends for the individual properties 
of the different atoms are not similar. It can be noticed here that the RPA values for the polarizabilities of the inert gas atoms are slightly larger than the DHF values and the CCSD values increase
further and are close to the experimental value for $^{199}$Xe. The experimental value of this quantity for $^{223}$Rn is not available but from the similarities in the
trends of electron correlation effects in both the inert gas atoms, we expect that our estimated $\alpha_d$ of this atom is accurate and will serve as a benchmark for its 
measurement. In contrast, this trend is different for the polarizabilities of the other atoms where the outer two orbitals can be treated as valence orbitals and the electron correlation effects between 
them are very strong. The RPA values are larger than those of their CCSD counterparts. It is interesting to point out that the CCSD method implicitly contains  the RPA effects. This highlights the very 
strong cancellations between the core-polarization effects, represented by RPA, and the non-core-polarization correlations to obtain the $\alpha_d$ values in these atoms. The CCSD 
values are also closer to the available experimental results. The correlation trends are found to be different for $\alpha_d$ in $^{199}$Hg  than they are for 
the other two-valence atoms $^{171}$Yb and $^{225}$Ra. For $^{199}$Hg, the RPA value is larger than that of its DHF counterpart, but in the case of the CCSD it is the opposite, implying that
non-core-polarization effects to be very strong. For the other two atoms, the CCSD results lie in between the DHF and RPA values. We have also given estimates of the  corrections due to the Breit interaction 
($\Delta$B) using RPA and found that they are very small. Compared to $\alpha_d$ values, the correlation trends in the $d_a$ values are found to be different. In these cases, the
RPA values are always larger. The DHF and CCSD values are very close for the inert gas atoms while in other atoms both the RPA and CCSD values are quite larger than the DHF values. 
Electron correlation effects contribute significantly in $^{171}$Yb and $^{225}$Ra than in the case of$^{199}$Hg. The Breit interaction also give very small contributions. Since the perturbations 
are odd-parity vector operators, the variations in the correlation trends are mainly due to the different radial behavior of these operators. The major contributions to the matrix elements of the 
electric dipole operator and the P,T-odd interaction Hamiltonians  come from regions far away and close to the nucleus respectively. For all the quantities, we also give the ``Final'' values by 
considering the CCSD results and estimating the uncertainties due to the Breit interaction, neglected triples and the truncated basis size effects. Most of the contributions come from both the 
triples, which is estimated using the operator defined by Eq. (\ref{eq30}), and basis-size effects.

 We also present results obtained from other methods in Table \ref{tab1}. Our RPA values agree well with the RPA values of Ref. \cite{dzuba09}. Our CCSD results do not match 
with those of other calculations as they take into account more physical effects. The PRCC and our CCSD methods are based on the same RCC formalism, but there is a difference in the 
implantation in the two cases. Unlike our approach, the normalization factor of the wave function appears explicitly in the PRCC method \cite{latha}. The CI$+$MBPT method 
takes the  $V^{N_c-2}$ potential in generating the orbitals for the  reference state while our DHF wave function is based on the $V^{N_c}$ potential. The CI$+$MBPT and MCDF methods are not
 size-extensive unlike the RCC method. Besides our RCC work, the only other method used for the inert gas atoms is the RPA.  

 To demonstrate how the correlations effects contribute through different RCC terms to our results, we give their values by classifying them into three terms as 
$\tilde{D}_{ob}T_1^{(1)}$, $T_1^{(0)\dagger}\tilde{D}_{ob}T_2^{(1)}$ and $T_2^{(0)\dagger}\tilde{D}_{ob}T_2^{(1)}$ along with their h.c. terms for the properties 
in Table \ref{tab2}.  $\tilde{D}_{ob}$ represents the effective one-body part  of the terms in the non-terminating series $\tilde{D}$ of Eq. (\ref{eqccx}) and it is determined self-consistently. By 
multiplying this effective one-body operator with the perturbed $T_1^{(1)}$, we get the most dominant contribution. This term includes all the RPA contributions (that also 
contains the lowest-order DHF results) and the other dominant non-core-polarization effects as the net singly excitations. The other two terms represent contributions 
purely due to the non-core-polarization effects. The contributions from the later two terms are very small in the evaluation of $d_a$, but  they 
contribute significantly to the polarizabilities. The differences between the RPA values from  $\tilde{D}_{ob}T_1^{(1)}$ contributions suggest the 
importance of the non-core-polarization correlation effects in these studies. Therefore, it is imperative to employ a suitable many-body method that can treat all possible 
correlations on equal footing for reliable calculations of the EDMs and the dipole polarizabilities of atoms that we have considered.

\section{Conclusion}
 
We have demonstrated that the RCC theory is capable  of determining the EDMs and the dipole polarizabilities of diamagnetic atoms to high accuracy. The importance of accounting for non-core-polarization effects  to all-orders through the RCC theory for these properties in atoms,
that are neglected in RPA have been highlighted. We have also shown that the electron correlation trends
for EDMs and polarizabilities differ even though operators corresponding to them are both odd-parity vector operators. This suggests that the electron correlation 
trends are strongly influenced by the radial behavior of the operators. Since the relativistic CCSD method is the most rigorous of all the methods we have employed to evaluate EDMs, it would be appropriate to combine the results based on it with those of the measured EDMs of diamagnetic atoms to obtain CP violating coupling constants in order to probe new physics beyond the standard model.

\section*{Acknowledgement}

B. K. S. acknowledges financial support from Chinese Academy of Science (CAS) through the PIFI fellowship under the project number 2017VMB0023 and partly by the TDP project of 
Physical Research Laboratory (PRL). Computations were carried out using Vikram-100 HPC cluster of PRL, Ahmedabad, India.


\end{document}